\documentclass[journal]{IEEEtran}
\usepackage{booktabs, multicol, multirow}
\usepackage{cite}

\ifCLASSINFOpdf
  \usepackage[pdftex]{graphicx}
   \graphicspath{{../pdf/}{../jpeg/}}
   \DeclareGraphicsExtensions{.pdf,.jpeg,.png}
\else
\fi
\usepackage{amsmath}
\usepackage{amssymb}
\usepackage{flexisym}
\usepackage{array}
\newcolumntype{C}[1]{>{\centering\let\newline\\\arraybackslash\hspace{0pt}}m{#1}}

\usepackage{color}
\definecolor{dblue}{rgb}{0,0,0.8}

\usepackage{xcolor}

\usepackage{subfig}
\usepackage{graphicx}
\usepackage{fixltx2e}

\usepackage{stfloats}
\usepackage{comment}

\usepackage{ifpdf}

\usepackage{algorithmicx}
\usepackage{times}
\usepackage{fancyhdr,graphicx,amsmath,amssymb}
\usepackage[ruled,vlined]{algorithm2e}
\usepackage{cite}
\include{pythonlisting}
\makeatletter
\def\BState{\State\hskip-\ALG@thistlm}
\makeatother

\usepackage{cite}
\usepackage{amsmath}
\usepackage{amssymb}
\usepackage{flexisym}
\usepackage{array}
\newcolumntype{C}[1]{>{\centering\let\newline\\\arraybackslash\hspace{0pt}}m{#1}}

\usepackage{color}
\definecolor{dblue}{rgb}{0,0,0.8}

\usepackage{xcolor}

\usepackage{multirow}

\usepackage{subfig}
\usepackage{graphicx}
\usepackage{fixltx2e}

\usepackage{stfloats}

\usepackage{subfig}

\usepackage[utf8]{inputenc}

\usepackage{amssymb}

\usepackage{nomencl}
\makenomenclature
\usepackage{etoolbox}
\usepackage{hyperref}

\usepackage{url}

\usepackage{mhchem}

\usepackage{nomencl}
\makenomenclature

\usepackage[justification=centering]{caption}

\usepackage{nomencl}
\makenomenclature

\usepackage{setspace}

\usepackage{hyperref}

\begin{document}
%



\title{Cyber-Physical Vulnerability Assessment of P2P Energy Exchanges in Active Distribution Networks}


\author{\IEEEauthorblockN{Hamed Haggi, \textit{Student Member, IEEE}, Wei Sun, \textit{Senior Member, IEEE}}\\
\IEEEauthorblockA{Department of Electrical and Computer Engineering, University of Central Florida, Orlando, FL, USA\\
Emails: hamed.haggi@knights.ucf.edu, sun@ucf.edu}

\thanks{This material is based upon work supported by the U.S. Department of Energy’s Office of Energy Efficiency and Renewable Energy (EERE) under the Solar Energy Technology Office (SETO) Award Number DE-EE0009339. \par}
}

\maketitle
\thispagestyle{plain}
\pagestyle{plain}
\begin{abstract}
Owing to the decreasing costs of distributed energy resources (DERs) as well as decarbonization policies, power systems are undergoing a modernization process. The large deployment of DERs together with internet of things (IoT) devices provide a platform for peer-to-peer (P2P) energy trading in active distribution networks. However, P2P energy trading with IoT devices have driven the grid more vulnerable to cyber-physical threats. To this end, in this paper, a resilience-oriented P2P energy exchange model is developed considering three phase unbalanced distribution systems. In addition, various scenarios for vulnerability assessment of P2P energy exchanges considering adverse prosumers and consumers, who provide false information regarding the price and quantity with the goal of maximum financial benefit and system operation disruption, are considered. Techno-economic survivability analysis against these attacks are investigated on a IEEE 13-node unbalanced distribution test system. Simulation results demonstrate that adverse peers can affect the physical operation of grid, maximize their benefits, and cause financial loss of other agents.  

\end{abstract}

\vspace{1.5mm}

%
\IEEEpeerreviewmaketitle

\section{Introduction}
\IEEEPARstart{D}{ecreasing} costs of distributed energy resources (DERs) and net-zero emission energy production policy are two preeminent factors that motivate utilities to deploy more DERs to enable deep decarbonization of energy production targets \cite{IEA}.  Since energy decarbonization cannot be achieved without high penetration of renewable energy sources, utilities should develop and invest in new business models by considering the price-making role of customer side DERs. However, the current market designs are not able to provide sufficient incentives for small-scale DER owners to share the energy with grid or neighbors. To this end, in order to capture the socio-economic benefits of customer side DERs, peer-to-peer (P2P) energy exchange platforms, including communication, business, and physical layers, are introduced to maximize the green energy harvesting, reducing the bills of customers, etc. However, The large number of peers (prosumers and consumers) equipped with internet of things (IoT) devices have driven the power grids more complex and vulnerable to cyber-physical threats such as natural disasters, cyber intrusions, etc. \cite{haggi2019review}.  Therefore, the motivation of this paper is to explore the potential cyber-physical threats in P2P markets and analyze the vulnerability of these threats on the real operation of power systems from economic and physical perspectives. \par

Research studies on P2P energy exchanges in the literature are broadly focused on the market designs with considering networks' physical constraints, network usage charges, communication layer, etc. using different approaches. For instance, in \cite{sampath2021peer,baroche2019exogenous,wang2020distributed} distributed optimization techniques are used for optimal operation of prosumers while addressing the privacy issues in P2P markets. An iterative mechanism is proposed in \cite{kim2020p2p} for peer and system-centric trading using direct negotiation through price adjustment process. To consider the role of consumers as price maker entities, double auction-based methods are widely used in P2P markets. For example, a continuous double auction-based P2P energy trading is proposed in \cite{guerrero2018decentralized} considering the network's physical constraints. An iterative double auction is used for localized P2P electricity trading in \cite{kang2017enabling} considering consortium blockchain technology. In \cite{haggi2021multi}, a hierarchical method based on multi-round double auction is proposed considering network charges. Cooperative and non-cooperative game theory-based P2P mechanisms are widely discussed in \cite{anoh2020energy,cui2020new,tushar2019grid} in which a unique and stable equilibrium can be achieved. More details regarding the P2P market designs and frameworks can be found in \cite{tushar2020peer}. \par

As mentioned above, a large number of prosumers and consumers in P2P designs are equipped with IoT devices to share required information with each other or system operators. The information exchange happen through the communication layer of P2P designs and transactive energy markets. However, the communication and physical layers of P2P-enabled distribution network models are in danger of cyber-physical attacks. For instance, various data manipulation and false data injection attacks in power systems are reviewed in \cite{liang2016review, dasgupta2021cyber}. A cooperative learning-based decentralized P2P energy trading market with considering cyber issues is proposed in \cite{Nguyen2021cooperative}. In \cite{barreto2020cyber}, a blockchain-based transactive energy market defense behavior against denial of service attacks is investigated. Cyber-physical security against data manipulation for transactive energy systems considering ventilation, heating, and air conditioning (HVAC) is analyzed in \cite{zhang2019cyber, zhang2019cyber2, krishnan2018cyber}. \par

P2P energy trading frameworks presented in \cite{sampath2021peer}-\cite{tushar2020peer} are mainly focused on various market mechanisms, while some of them are focused on modeling the physical network constraints and transaction charges in balanced distribution networks. However, the real distribution networks are phase unbalanced, and proper modelling efforts are required to integrate P2P models to unbalanced systems. Moreover, these research studies are only focused on the normal operation of system, and neglect to consider the emergency operation in the case of disruptions such as natural disasters, cyber attacks, etc. Research works presented in \cite{liang2016review}-\cite{krishnan2018cyber} review the potential cyber-physical threats in transactive energy markets. Only in \cite{zhang2019cyber, zhang2019cyber2, krishnan2018cyber} the impact of false data injection on HVAC systems is investigated. There is no modelling effort for resilience-oriented P2P interactions in unbalanced network considering the physical constraints such as voltage, line loading, congestion, etc. Motivated by the aforementioned challenges, this paper proposes a resilience-oriented framework for P2P energy exchanges in unbalanced active distribution networks focusing on both normal and emergency operation modes. In addition, to analyze the vulnerability against cyber-physical attacks, adverse prosumers and consumers who provide false information regarding their surplus renewable energy, demand, and offered price are considered, which results in financial losses of other agents, physical constraint violation, load curtailment, etc. \par

The rest of the paper is organized as follows. The proposed P2P framework with cyber-physical threats is explained in Section \ref{P2PThreats}. Problem formulation and simulation results and analysis are presented in Sections \ref{formulation} and \ref{results}, respectively. Finally, Section \ref{conclusion} concludes the paper and presents the future directions of this research study.

\begin{figure*}
\centering
\footnotesize
\captionsetup{justification=raggedright,singlelinecheck=false,font={footnotesize}}
	\includegraphics[width =5.7in,height=2.8in]{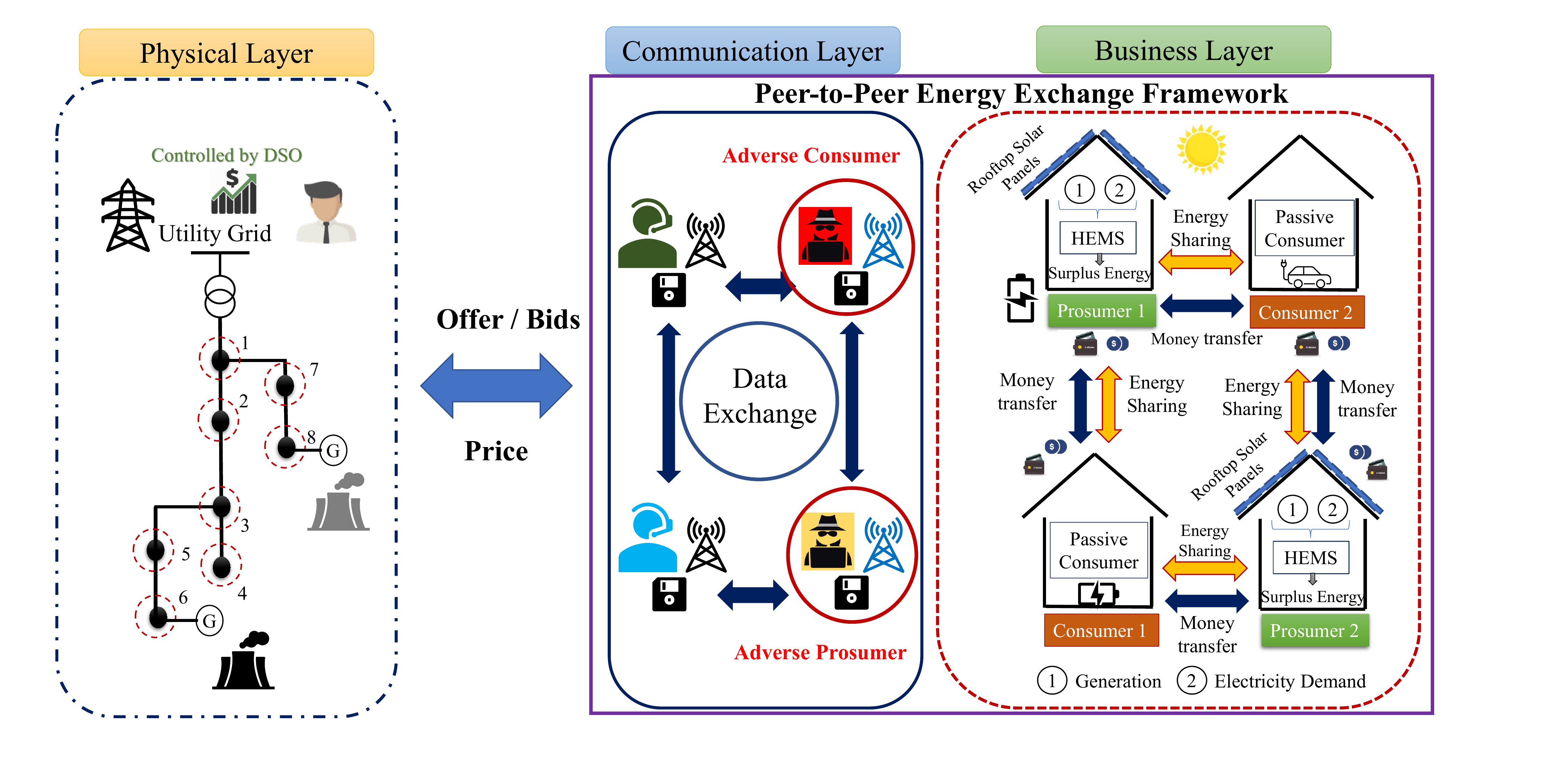}
	\caption{Peer-to-Peer enabled distribution system operation with business, physical, and communication layers.}
    \label{framework}
\end{figure*}

\section{Proposed P2P Energy Exchange Framework with Cyber-Physical Threats} \label{P2PThreats}

Fig. \ref{framework} shows the P2P-enabled distribution system operation including physical, communication, and business layers for agents (including prosumers and consumers that might be equipped with home energy management systems, batteries, etc.). During the time when the prosumers have surplus energy, they can share required information including price and quantity through communication layer with consumers to exchange energy based on business layer rules, and increase the monetary benefits while addressing the physical constraints of power grid. However, P2P markets can face with preeminent challenges such as cyber-physical threats due to disruptions in any layers. Natural disasters such as hurricanes, floods, wildfire, etc. and physical attacks on assets in distribution system can impose a threat on the grid-level operation \cite{haggi2019review, mehrdad2018cyber}. Moreover, P2P markets create opportunities for cyber attacks at different layers especially communication layer due to large number of IoT devices. Attacks on communication channels (e.g., false data injection, load redistribution attacks, etc.), attacks through adverse users (providing false information to operators and agents), and physical attacks (e.g., on smart metering or distribution lines) are potential threats that P2P energy sharing mechanisms can face \cite{dasgupta2021cyber}. This paper is focused on prosumers' and consumers' intentional attack as an attacker by providing false information regarding their surplus energy, demand, or even the price of selling or purchasing energy. It should be mentioned that coordinated attack among prosumers and consumers as adverse agents can significantly affect the normal operation of system from economic and technical perspectives.

\section{Problem Formulation}\label{formulation}
The proposed resilience-oriented P2P formulation which is integrated to unbalanced distribution network is presented in this section. The distribution network is represented as a graph ($\mathcal{N}$,$\mathcal{L}$), where $\mathcal{N}$ and $\mathcal{L}$ are the set of nodes and lines, respectively. Additionally, the line impedance is expressed as $Z_l = R_l + jX_l$. The set of prosumers' and consumers' nodes are defined as $\mathcal{N}^p$ (indexed by $i$) and $\mathcal{N}^c$ (indexed by $j$), correspondingly in which $\mathcal{N}^p \cup  \mathcal{N}^c = \mathcal{N}$. Moreover, to make the formulation general, we consider the time index in the formulation and $\mathcal{T}$ represents the set of time steps indexed by $t$. \par

\subsection{P2P Energy Exchange Formulation}
The P2P energy trading formulation with the goal of minimizing the total cost of peers (maximizing social welfare) is presented in equations (\ref{P2POF1})-(\ref{P2P10}):

\begin{equation}\label{P2POF1}
\text{min.} \; OC^{P2P} = \sum_{t \in T} \Bigg\{ \sum_{i \in N^p} C_i(P_{t,i}^p) - \sum_{j \in N^c} U_j(dem_{t,j}^c) \Bigg\}
\end{equation}
\begin{equation}\label{P2P2}
P_{nm}(t,i,j)\;+\;P_{mn}(t,j,i) = 0, \quad \forall (i,j) \in (N^p,N^c)
\end{equation}
\begin{equation}\label{P2P3}
    P_{nm}(t,i,j) \ge 0, \quad \forall (i,j) \in (N^p,N^c)
\end{equation}
\begin{equation}\label{P2P4}
    P_{mn}(t,j,i) \le 0, \quad \forall (i,j) \in (N^p,N^c)
\end{equation}
\begin{equation}\label{P2P5}
    P^p_{t,i} = \sum_{j} P_{nm}(t,i,j), \quad \forall (i,j) \in (N^p,N^c)
\end{equation}
\begin{equation}\label{P2P6}
    dem^c_{t,j} = \sum_{i} P_{mn}(t,j,i), \quad \forall (i,j) \in (N^p,N^c)
\end{equation}
\begin{equation}\label{P2P7}
    P^{p,min} \le P^p_{t,i} \le P^{p,max}, \quad \forall i \in N^p
\end{equation}
\begin{equation}\label{P2P8}
    Q^{p,min} \le Q^p_{t,i} \le Q^{p,max}, \quad \forall i \in N^p
\end{equation}
\begin{equation}\label{P2P9}
    P^{Load,min} \le dem^c_{t,j} \le P^{Load,max}, \quad \forall j \in N^c
\end{equation}
\begin{equation}\label{P2P10}
(P^p_{t,i})^2 + (Q^p_{t,i})^2 \le (S^{inv})^2, \quad \forall i \in N^p
\end{equation}
where the objective function in (\ref{P2POF1}) includes the prosumers cost and utility of consumers (linear cost curves). $P_{t,i}^p$ and $dem_{t,j}^c$ denote to total traded power of prosumers and total received power by consumers, respectively. Equations (\ref{P2P2})-(\ref{P2P4}) illustrate the P2P energy exchange among prosumers and consumers in which the sold power by prosumer, $P_{nm}$, should be equal with the power purchased by consumer $P_{mn}$. Equation (\ref{P2P5}) and (\ref{P2P6}) express the total sold power from prosumer to consumers, and total purchased power by consumer from other prosumers, respectively. Active and reactive power (reactive ower is defined as $Q^{p}_{t,i}$) constraints of prosumers, and general form of elastic demand for all agents are expressed in (\ref{P2P7})-(\ref{P2P9}). However, in this paper it is assumed that the agents' demand is considered as inelastic. Therefore, the upper and lower limits are the same. The inverter capacity constraints of prosumers (e.g. rooftop solar owners) is shown in (\ref{P2P10}) which is defined as $S^{inv}$. 
\subsection{Unbalanced Grid Operation Formulation}
To address the business and physical network constraints, the aforementioned P2P model is integrated to unbalanced convex branch flow \cite{farivar2013branch} model, as shown in (\ref{DistflowOF})-(\ref{shedreactive}):
\begin{equation}\label{DistflowOF}
\text{min.} \; OC^{Grid} = \sum_{t \in T}  \Bigg\{ \sum_{i \in N^g} C_i(P_{t,i}^{UG}) +  \sum_{i \in N} C_i(P_{t,i}^{Shd}) \Bigg\}
\end{equation}
\begin{equation}\label{activebalance}
f^p_{t,i} = P^{Load}_{t,i} + \sum_{j\rightarrow i} f^p_{t,j} + R_{l}.a_{t,l} - P^{UG}_{t,i} - P^p_{t,i} - P^{Shd}_{t,i}    
\end{equation}
\begin{equation}\label{reactivebalance}
f^q_{t,i} = Q^{Load}_{t,i} + \sum_{j\rightarrow i} f^q_{t,j} + X_{l}.a_{t,l} - Q^{UG}_{t,i} - Q^p_{t,i} - Q^{Shd}_{t,i}       
\end{equation}
\begin{equation}\label{linelimit1}
(f^p_{t,l})^2 + (f^q_{t,l})^2 \le (S^{l})^2
\end{equation}
\begin{equation}\label{linelimit2}
(f^p_{t,i} -  R_{l}.a_{t,l} )^2 + (f^q_{t,i}-  X_{l}.a_{t,l} )^2 \le (S^{l})^2
\end{equation}
\begin{equation}\label{SOCPconvex}
\begin{Vmatrix}
2 f^p_{t,i}\\ 
2 f^q_{t,i}\\ 
a_{t,l} - v_{t,i}
\end{Vmatrix}_2 \leq a_{t,l} + v_{t,i}
\end{equation}
\begin{equation}\label{voltage}
v_{t,i} = v_{t,j} - 2(\widetilde{R_{l}}\;.\;f^p_{t,i} + \widetilde{X_{l}}\;.\;f^q_{t,i}) + \widetilde{Z_{l}}\;.\; a_{t,l}
\end{equation}
\begin{equation}\label{voltagelimit}
(V^{min})^2 \le v_{t,i} \le (V^{max})^2
\end{equation}
\begin{equation}\label{currentlimit}
(I^{min})^2 \le a_{t,l} \le (I^{max})^2
\end{equation}
\begin{equation}\label{shedcost}
C_i(P_{t,i}^{Shd}) = VOLL .\;P^{Shd}_{t,i}
\end{equation}
\begin{equation}\label{shedlimit}
0 \le P^{Shd}_{t,i} \le P^{Load}_{t,i}
\end{equation}
\begin{equation}\label{shedreactive}
Q^{Shd}_{t,i} = P^{Shd}_{t,i}\;.\;\frac{Q^{Load}_{t,i}}{P^{Load}_{t,i}}
\end{equation}
where the objective is to minimize the total purchased power from upper grid, $P^{UG}_{t,i} \in \mathbb{R}^{3 \times 1}$, and load curtailment, $P^{Shd}_{t,i} \in \mathbb{R}^{3 \times 1}$, due to cyber-physical threats. Let us denote $f_{t,i}^{p/q} \in \mathbb{R}^{3 \times 1}$, $R_l$, $X_l$, $a_{t,i} \in \mathbb{R}^{3 \times 1}$, $Q^{Shd}_{t,i} \in \mathbb{R}^{3 \times 1} $, $Q^{UG}_{t,i} \in \mathbb{R}^{3 \times 1}$ the active/reactive power flow of distribution lines, line resistant, line reactance, squared current flow of the lines, reactive load curtailment, and reactive power from upper grid. In addition, $S^l$, $v_{t,i} = \left [ \left | v_{t,i}^a \right |, \left | v_{t,i}^b \right |, \left | v_{t,i}^c \right |\right ]^T$, $I^{min/max}$, $V^{min/max}$ denote line capacity, squared voltage magnitude, and minimum and maximum limits for the current and voltage magnitudes. Considering the definitions, (\ref{activebalance}) and (\ref{reactivebalance}) refer to active and reactive power balance of distribution network including the peers demand and surplus energy. The apparent power flow limits and relaxation of second order cone constraint are expressed in (\ref{linelimit1}), (\ref{linelimit2}), and (\ref{SOCPconvex}). Voltage drop constraints and its maximum and minimum limits as well as current limits are expressed in (\ref{voltage})-(\ref{currentlimit}). More information regarding the angle relaxation and balanced voltage, and also distributional locational marginal price (DLMP) can be found in \cite{nejad2019distributed} and \cite{papavasiliou2017analysis}, respectively. Finally, load shedding cost based on the value of loss of load (VOLL) and active and reactive load curtailment constraints are expressed in (\ref{shedcost})-(\ref{shedreactive}). \par

Considering $\odot$ as element wise product, the following equations are used to calculate $\widetilde{R_{l}}$, $\widetilde{X_{l}}$, and $\widetilde{Z_{l}}$ \cite{nejad2019distributed}:
\begin{equation}\label{Rhat}
\widetilde{R_{l}} = Re\left \{ \alpha \;  \odot \; R_l \right \}  - Img \left \{ \alpha \; \odot \; X_l \right \}
\end{equation}
\begin{equation}\label{Xhat}
\widetilde{X_{l}} = Re\left \{ \alpha \;  \odot \; X_l \right \}  + Img \left \{ \alpha \; \odot \; R_l \right \}
\end{equation}
\begin{equation}\label{Zhat}
\widetilde{Z_{l}} = \left | Z_{l} \right | \odot \left | Z_{l} \right |
\end{equation}
\begin{equation}\label{alphaconvert}
\alpha =  \begin{bmatrix}
1 & e^{-j2\pi/3} & e^{j2\pi/3}\\ 
 e^{j2\pi/3} &  1& e^{-j2\pi/3}\\ 
 e^{-j2\pi/3}& e^{j2\pi/3} & 1
\end{bmatrix}
\end{equation}

The final co-optimization between P2P model and distribution system is expressed in (\ref{FinalP2P}).
\begin{equation}\label{FinalP2P}
\begin{split}
\text{min.}\; [\text{ $OC^{P2P}$ + $OC^{Grid} $ }]  \qquad \qquad \qquad \quad\\
 \text{s.t} \qquad \qquad \qquad \qquad \qquad \qquad \qquad  \\
 \text{Equations (\ref{P2P2})-(\ref{P2P10}), (\ref{activebalance})-(\ref{alphaconvert})} \qquad \qquad
\end{split}
\end{equation}

\section{Simulation Results and Analysis}\label{results}

\begin{figure}
\centering
\footnotesize
\captionsetup{justification=raggedright,singlelinecheck=false,font={footnotesize}}
	\includegraphics[scale=0.7]{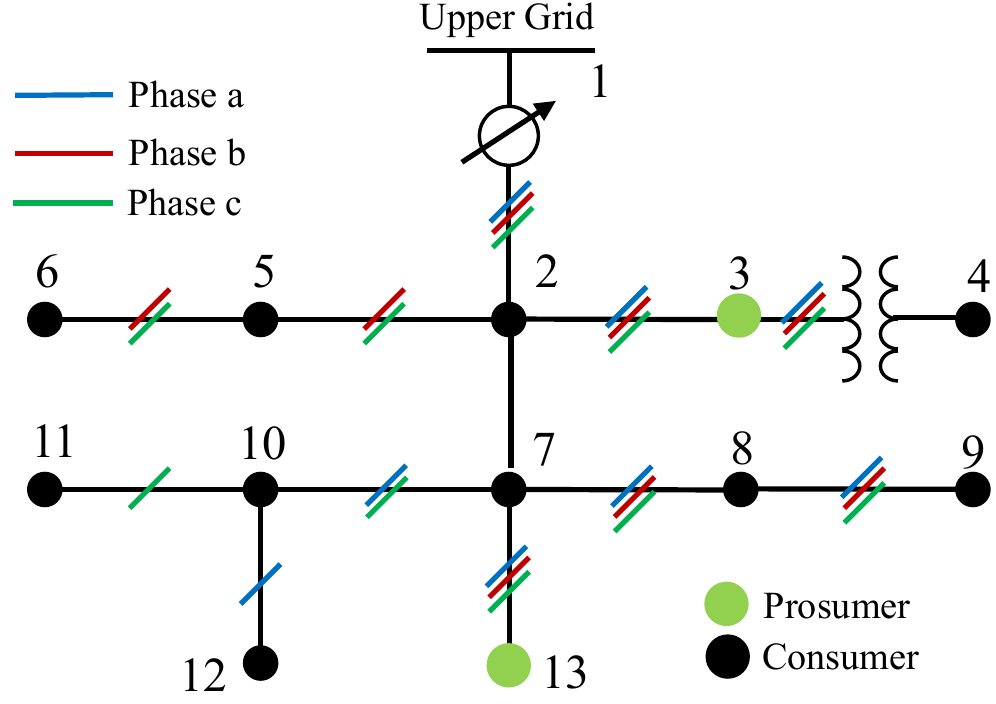}
	\caption{Modified IEEE 13-node distribution test system.}
    \label{Testsystem}
\end{figure}

The proposed resilience-oriented framework is validated by testing on the modified IEEE-13 node unbalanced distribution system shown in Fig. \ref{Testsystem}. More details regarding the test system can be found in \cite{Feeders}. Nodes 3 and 13 are considered as prosumers (with three-phase connection) with maximum capacity of 650 KVA and offering price of \$35/MWh and \$20/MWh, respectively. Prosumer at node 13 and consumer at node 7 are considered as adverse agents (attackers who provide false information regarding the price and quantity). During the post attack scenario, prosumer at node 13 intentionally increases the price of energy sharing to \$45/MWh and consumer at node 7 reports its demand 25\% higher to violate the regular operation of system. Two scenarios for simulation results are considered; 1) Techno-economic survivability analysis before the attack of adverse agents, 2) Post-attack survivability assessment. 

\subsection{Pre- and Post-attack Financial Analysis of Prosumers and Consumers}
\begin{figure}
\centering
\footnotesize
\captionsetup{justification=raggedright,singlelinecheck=false,font={footnotesize}}
	\includegraphics[scale=0.7]{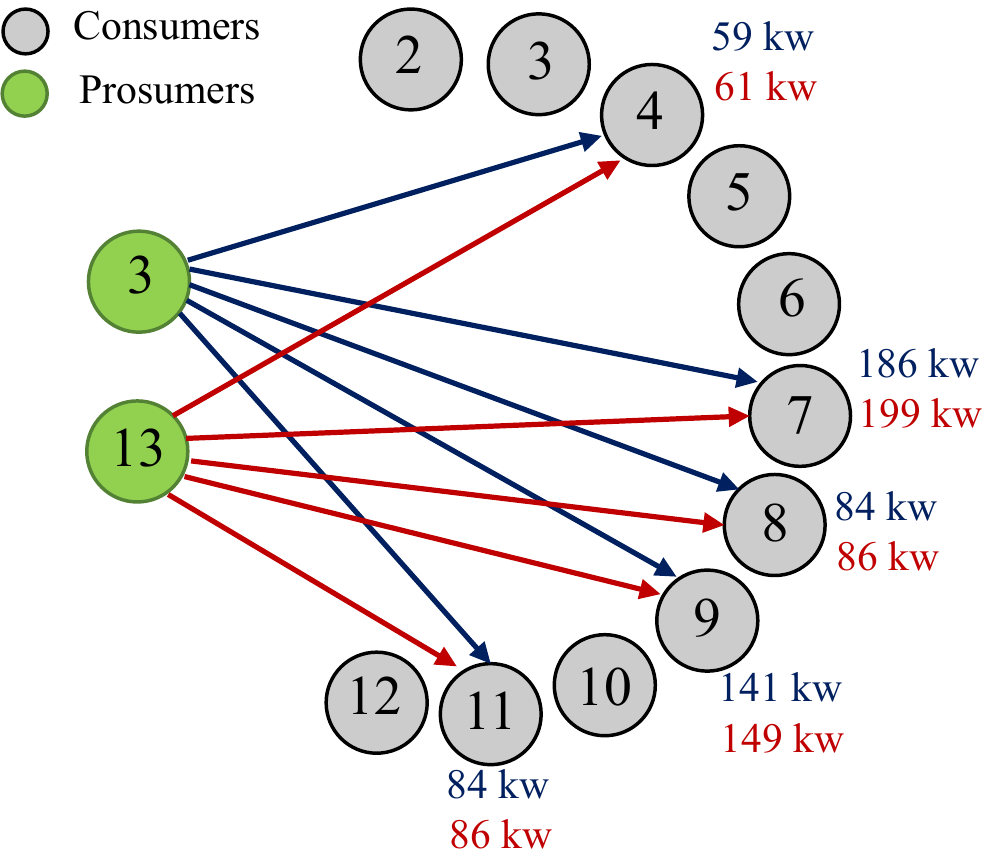}
	\caption{Pre-attack P2P matchings among prosumers and consumers (phase c).}
    \label{peermatching}
\end{figure}

\begin{figure}
\footnotesize
\captionsetup{justification=raggedright,singlelinecheck=false,font={footnotesize}}
 \subfloat[Three phase DLMP before attack] {\includegraphics[width=1.76in,height=1.6in]{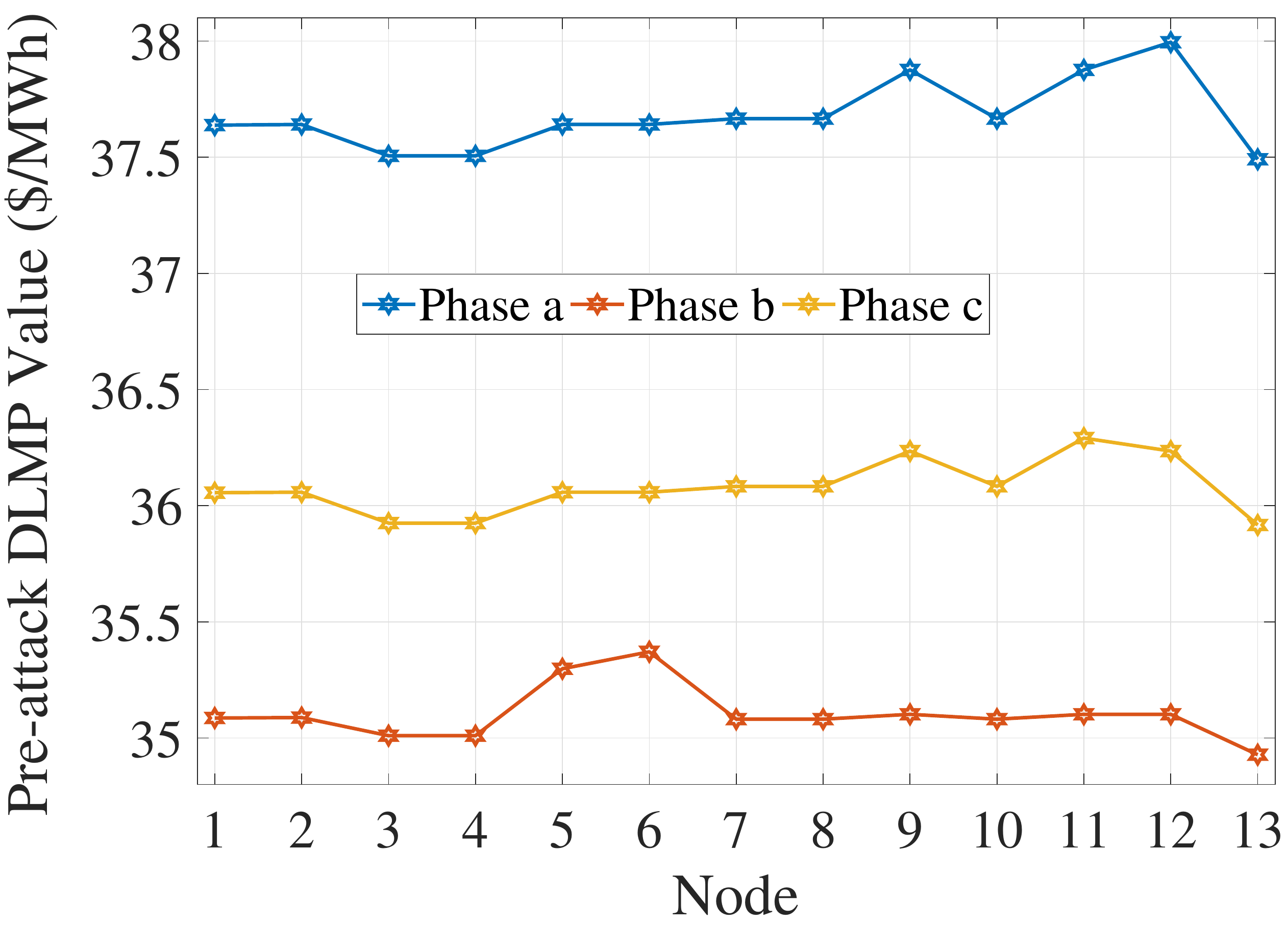}\label{VAp2p}}
 \subfloat[Three phase DLMP after attack] {\includegraphics[width=1.72in,height=1.625in]{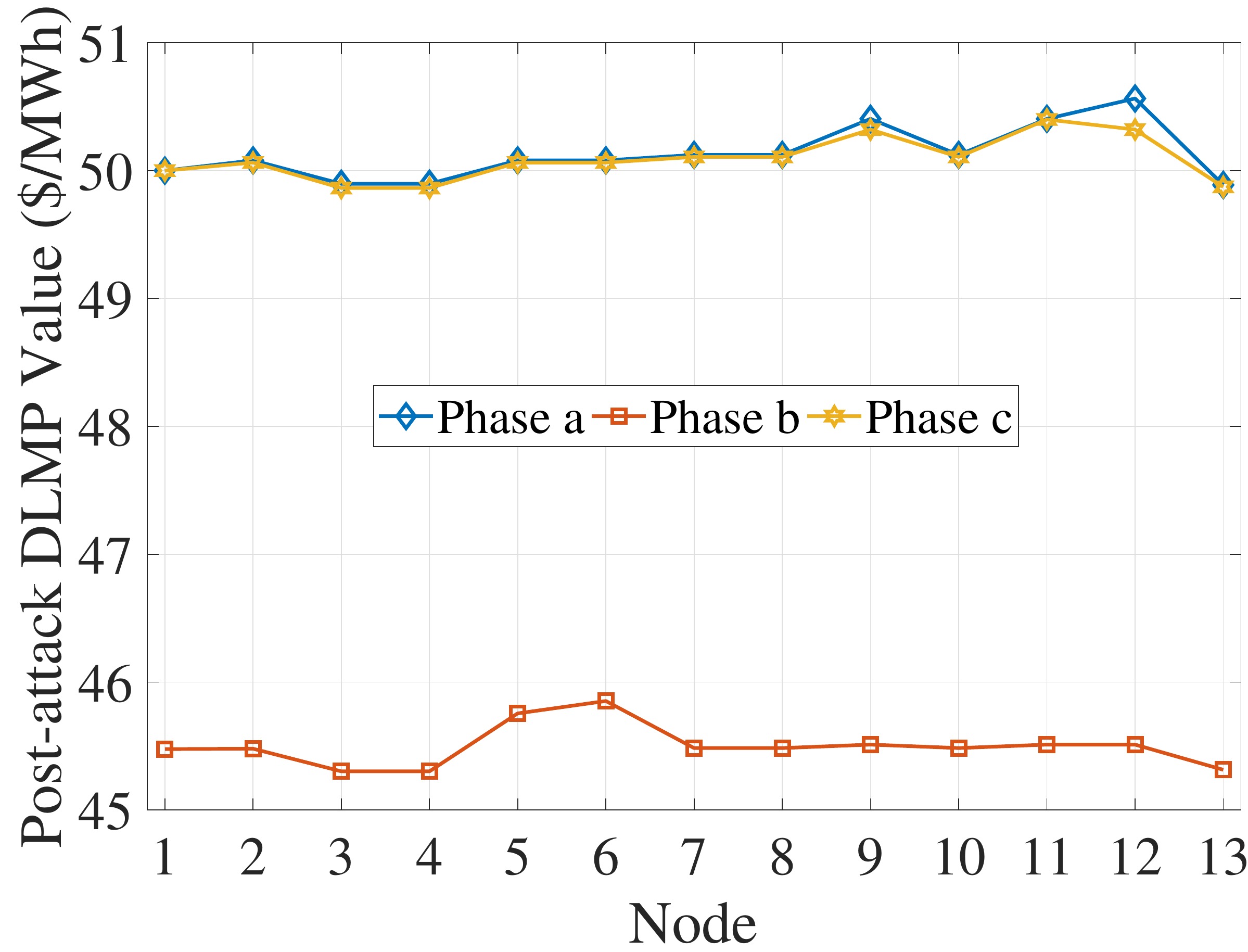}\label{VNO}}
 \caption{IEEE 13-node DLMP values with and without coordinated attack of adverse prosumer and consumer.}
	\label{DLMP}
\end{figure}

The normal operation results of P2P energy exchanges for phase c of 13-node system is shown in Fig. \ref{peermatching}, in which two prosumers exchange power with consumers distributed across the grid. Due to no demand in nodes 2, 3, 5, 6, 10, and 12 \cite{Feeders}, there is not any matchings with prosumers in Fig. \ref{peermatching}. Any prosumer can exchange energy with multiple consumers at grid and vice versa. For instance, considering the phase c, consumer located at node 7 receives 385 kw from both prosumers located at nodes 3 and 13. In addition, DLMP for each phase is calculated and presented in Fig. \ref{DLMP}a. Before attack, prosumers share correct information with system operator, and the DLMP values are in the range of [\$35/MWh-\$38/MWh] depending on the location of nodes and voltage regulation, power loss , costs and electricity demand from consumers. During the coordinated attack, prosumer at node 13 and consumer at node 7 intentionally provide false information regarding the offering price and quantity with the goal of maximizing benefit and disruption in normal operation of system. Therefore, the DLMP values are shifted to the range of [\$45/MWh-\$51/MWh]. In this attack, consumer at node 7 increases the demand by 25\%, and the DLMP value at phase b significantly increases due to purchasing power from substation node with price of \$50/MWh, which results in more financial losses of consumers compared to phase c. \par

\begin{figure}
\centering
\footnotesize
\captionsetup{justification=raggedright,singlelinecheck=false,font={footnotesize}}
	\includegraphics[width=3.5in]{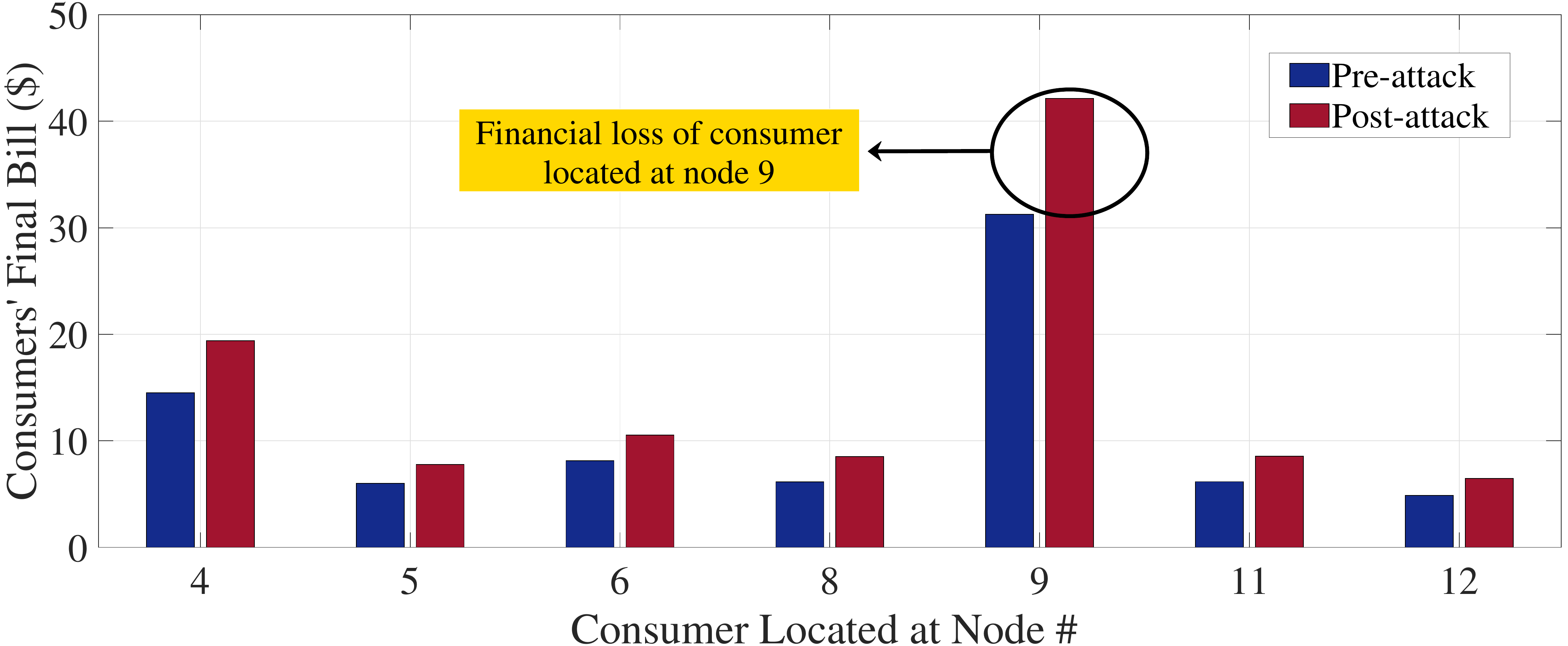}
	\caption{Financial loss of affected consumers before and after attack.}
    \label{financial_consumer}
\end{figure}

\begin{figure}
\centering
\footnotesize
\captionsetup{justification=raggedright,singlelinecheck=false,font={footnotesize}}
	\includegraphics[width=3.5in]{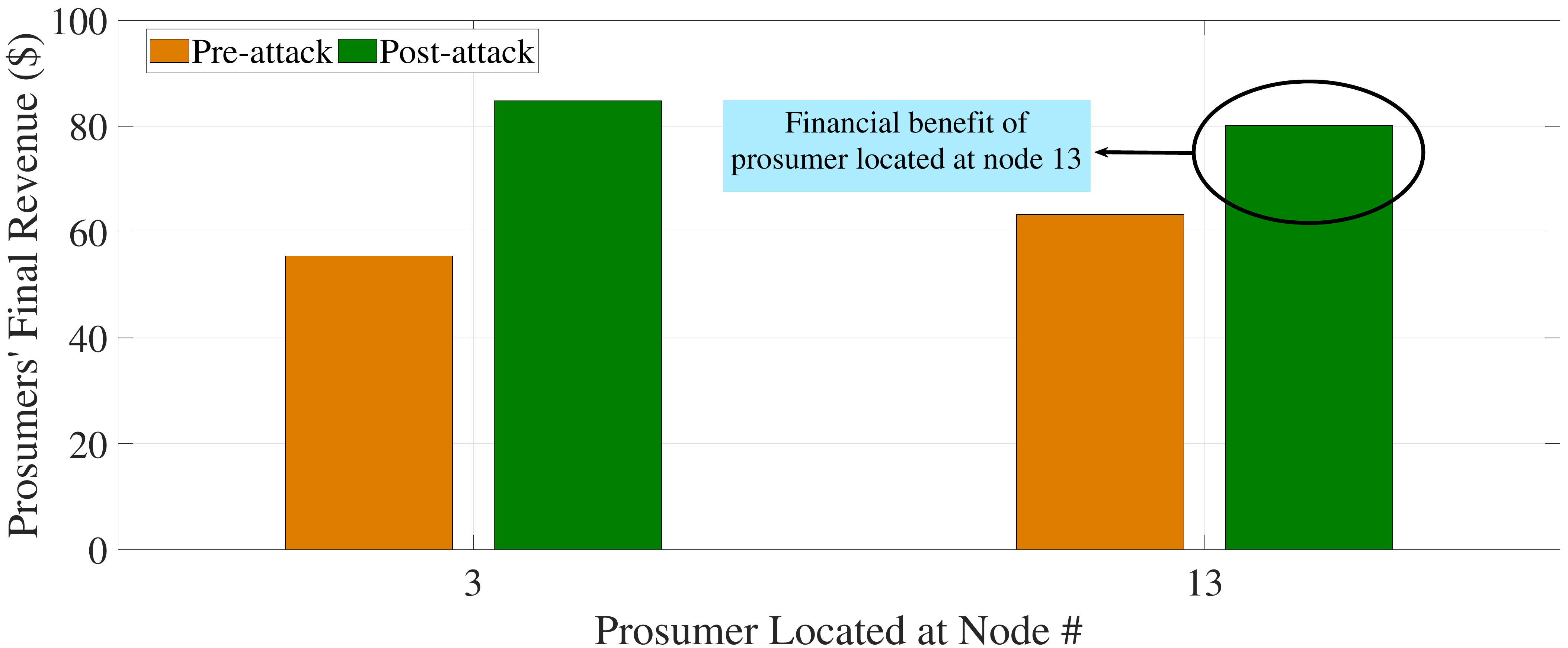}
	\caption{Financial benefit of adverse  prosumers before and after attack.}
    \label{financial_prosumer}
\end{figure}

The final bills of consumers and prosumers for both pre- and post-attack scenarios are calculated based on \cite{haggi2021multi}, as presented in Fig. \ref{financial_consumer} and Fig. \ref{financial_prosumer}. Due to the increase in DLMP in post-attack scenario, the final consumers' bill are increased too. For instance, the final bill of consumer 9 is increased by 35\% in the post-attack scenario, and the financial loss is clearly shown in Fig. \ref{financial_consumer}. It should be mentioned that node 9 has the highest electricity consumption among the affected consumers and that's why its loss is more than other consumers. Accordingly, the prosumers final revenue is increased in post-attack scenario, as shown in Fig. \ref{financial_prosumer}. It should also be noted that the total operation cost before and after attack are \$88.6 and \$144.5, respectively. 

\subsection{Results for Physical Constraints of Distribution Network before and after Attack}
Besides financial issues that adverse users can create for business layer of P2P market, they also have the capability of affecting the regular operation of networks. For instance, voltage magnitudes for three phases in both pre- and post-attack scenarios are presented in Fig. \ref{voltage}. It can be seen that adverse users can change the optimum voltage set points by providing false information, which also change the power flow of lines, additional costs, etc. However, since voltage is still within the range, it is hard for system operator to detect this attack. Additionally, prosumer located at node 13 can intentionally damage the line connecting node 2 to 7 (results in line outage) to increase the benefit by selling power to neighboring nodes with a very high price. For instance, if the line between nodes 2 to 7 is attacked to be out of service, the DLMP is increased in nodes 7 to 13 and as a consequence, prosumer 13 (as the only energy provider) can sell all the available power with a very high price such as $VOLL$ and maximize the benefit. Furthermore, this attack can results in 1.8 MW total curtailment in load.

\begin{figure}
\centering
\footnotesize
\captionsetup{justification=raggedright,singlelinecheck=false,font={footnotesize}}
	\includegraphics[width=3.5in]{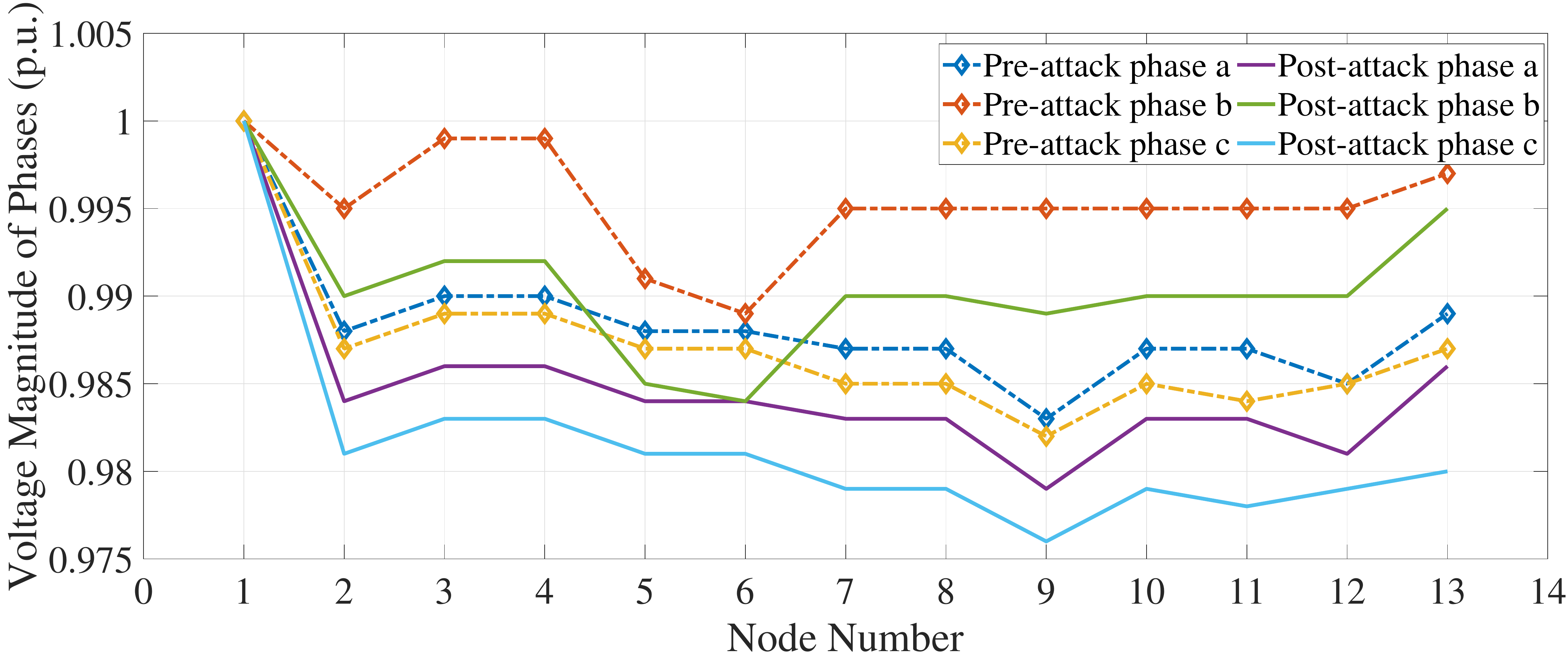}
	\caption{Voltage magnitudes before and after attack.}
    \label{voltage}
\end{figure}

\section{Conclusions}\label{conclusion}
This paper proposed a resilience-oriented framework for P2P energy exchanges in unbalanced active distribution networks. The problem is formulated to analyze the normal and emergency operation of P2P integrated distribution network operation in the case of cyber-physical threats. Besides analyzing the potential threats in P2P markets, adverse prosumers' and consumers' behaviour in providing false information regarding the offered price and quantity is considered. Additionally, the impact of this false data provision on market clearing price and physical constraints of network such as voltage is investigated. Simulation results on the IEEE 13-node test feeder demonstrate the effectiveness of the proposed model. The future direction of this work is to provide a defender-attacker-defender P2P model to guarantee a safe and reliable operation against cyber-physical attacks, such as false data injection, load redistribution attacks, etc.

\setstretch{0.95}

\bibliographystyle{IEEEtran}
\bibliography{mybib}

\end{document}